\documentclass[aps,preprint,prd,showpacs,nofootinbib]{revtex4}

\usepackage{amsmath}
\usepackage{graphicx}
\usepackage{dcolumn}
\usepackage{bm}
\usepackage{amssymb}
\usepackage{latexsym}
\usepackage{simplewick}
\usepackage{slashed,graphicx,color,amsmath,amssymb,simplewick}

\begin{document}

\title{Gravitational waves induced by spinor fields}

\author{Kaixi Feng$^1$\footnote{Email: fengkaixi10@mails.ucas.ac.cn}, Yun-Song Piao$^{1,2}$\footnote{Email: yspiao@ucas.ac.cn} }

\affiliation{$1$ School of Physics, University of Chinese Academy
of Sciences, Beijing 100049, China \vspace{1mm},\\ $2$ State Key Laboratory of Theoretical Physics, Institute of Theoretical Physics, \\
Chinese Academy of Sciences, P.O. Box 2735, Beijing 100190, China
\vspace{1mm}}

\begin{abstract}

In realistic model-building, spinor fields with various masses
are present. During inflation, spinor field may induce
gravitational waves as a second order effect. In this paper, we
calculate the contribution of single massive spinor field to the
power spectrum of primordial gravitational wave by using retarded
Green propagator. We find that the correction is scale-invariant
and of order $H^4/M_P^4$ for arbitrary spinor mass $m_{\psi}$.
Additionally, we also observe that when $m_\psi \gtrsim H$, the
dependence of correction on $m_\psi/H$ is nontrivial.

\end{abstract}
\maketitle

\section{Introduction}

General relativity strongly suggests the existence of
gravitational waves. Through the study of gravitational waves, we
might get rich information about new physics in the very early
universe directly since there may be footprints left in it if the
corresponding processes happened energetically. Recently, besides
B-mode polarization of CMB
\cite{Baumann:2008aq},\cite{Ade:2014xna}, gravitational waves
searching are proposed or underway on smaller scale by
interferometers like VIRGO, DECIGO and LISA, which make the
studies relevant with gravitational waves acquire
increasing attentions.

The primordial gravitational wave may be produced from quantum
fluctuations of vacuum during inflation
\cite{Grishchuk:1975ty},\cite{Starobinsky:1979ty},\cite{Rubakov:1982ty},
which is scale-invariant and is regarded as a smoking gun for
inflation. The primordial gravitational waves may record some
physical information during inflation, which can hardly be encoded
by scalar perturbation, e.g. a step-like variation of primordial
gravitational waves speed will result in the oscillating tensor
spectrum \cite{Cai:2015dta}. Thus, every possible way of affecting
gravitational waves spectrum is significant and worth
studying. There are also other mechanisms producing gravitational
waves in the early universe, e.g. cosmic defeats
\cite{Damour:2000wa, Figueroa:2012kw}, bubble collisions
\cite{Kosowsky:1991ua, Kamionkowski:1993fg}.




Back in the very early stage of universe, there must be numerous
fields besides the inflaton field. In linear order, the scalar
fields other than inflaton can only generate isocurvature
perturbations, see \cite{Chen:2009zp},\cite{Chen:2012ge} for the
effect of massive isocurvaton on the density perturbations, see
also \cite{Kehagias:2015jha},\cite{Arkani-Hamed:2015bza},\cite{Dimastrogiovanni:2015pla}.
When the action is expanded to third order in
perturbations, one may see that the scalar fields can
source gravity at quadratic order. This motivates the mechanism of
inducing gravitational waves by introducing a spectator scalar
field \cite{Biagetti:2013kwa}, see also
\cite{Biagetti:2014asa},\cite{Fujita:2014oba}.

Spinor fields with various masses should also be present in
early universe, as a consequence of realistic model-building.
During inflation, these spinor fields look like spectators, i.e.
they do not affect the inflationary dynamics at neither the
background nor the linear order fluctuations level.
Thus, similar to scalar spectators, they may also source
gravity at quadratic order. In addition, there are also some
studies about gravitational waves sourced by production of
particles see e.g.\cite{Cook:2011hg},\cite{Barnaby:2012xt}, by
production of fermion see \cite{Chung:1999ve, Figueroa:2013vif}.




There are mainly two kinds of methods of calculating
the corrections to the power spectrum. One is loop correction
with in-in formalism \cite{Weinberg:2005vy}. The loop correction
of a Dirac field to the scalar power spectrum was studied
comprehensively in \cite{Chaicherdsakul:2006ui}. In
Ref.\cite{Feng:2012jm}, we have calculated the loop correction of
Dirac field to the gravitational wave power spectrum. The other is
inducing gravitational waves as a second order effect,
using retarded Green propagator, and regarding the spectator
fields as sources.


In this paper, we will calculate the contribution of massive
spinor field during inflation to the gravitational wave power
spectrum by using retarded Green propagator. We find that the
correction is scale-invariant and of order $H^4/M_P^4$ for
arbitrary mass $m_{\psi}$. When $m_\psi \gtrsim H$, the dependence
of correction on $m_\psi/H$ is nontrivial. In Section II we give a
brief introduction to the background dynamics of spinor field in
inflationary cosmology. In Section III, we calculate the
primordial gravitational wave spectrum induced by spinor field and
give our result. The conclusion is given in Section IV. In Appendix A,
we show the calculation process of spin sum for massive fermions, which
is necessary for our computation of the massive spinor field's contribution
to the two point correlation function of tensor.

\section{ Dynamics of Dirac field}

In Ref.\cite{Feng:2012jm}, we have described in detail the
dynamics of a spinor field which is minimally coupled to Einstein
gravity, see also Refs. \cite{Weinberg, BirrellDavies} for
detailed introduction and \cite{Cai:2008gk} for the cosmology with
spinor fields. In this section, we will briefly review it.

Taking into account the spatial dependence of the Dirac field, the
equations of motion for it is given by
\begin{eqnarray}
\label{EoMpsi}
 i\gamma^0(\dot\psi+\frac{3}{2}H\psi)-i\gamma^i\frac{\partial_i}{a}\psi -m_\psi \psi &=&0~,
\end{eqnarray}
where $H$ is the Hubble parameter during inflation. As usual, the
Fourier transformation of $ \psi(\mathbf{x},t)$ is
\begin{eqnarray}
 \psi(\mathbf{x},t) = \int\mathrm{d}^3p~e^{i\mathbf{p}\cdot\mathbf{x}}\sum_s \left[\alpha_{\mathbf{p},s} X_{\mathbf{p},s}(t) + \beta^\dag_{-\mathbf{p},s} W_{-\mathbf{p},s}(t) \right]~,
\end{eqnarray}
where $s=\pm{1}/{2}$ is the spin of the Dirac field, and also the
annihilation operators of the Dirac field satisfies the
anti-commutative relation.


The Fourier modes $X_{\mathbf{p},s}(t)$, $W_{\mathbf{p},s}(t)$
satisfy the equations of motion shown in Eq.(\ref{EoMpsi}). Solving
the equation of motion in comoving time coordinate $\tau\equiv\int
dt/a$, we obtain
\begin{eqnarray}
\label{sol_psi}
 X_{k,\pm}(\tau) &=& \frac{i\sqrt{-\pi k\tau}}{2(2\pi a)^{3/2}} e^{\pm{\pi r}} H^{(1)}_{{1}/{2}\mp ir}(-k\tau)
 ~,
\end{eqnarray}
where $H^{(1)}$ is the first kind of the Hankel function and
$r=m_{\psi}/H$, and also we have made use of Bunch-Davies vacuum
as initial condition to determine the coefficient of the Hankel
function.

\section{Induced Gravitational Wave}

In Einstein gravity, the quadratic action of tensor perturbation
and the interaction between it and the Dirac field are
\begin{eqnarray}
S^{(2)}=\int d^4x {aM_P^2\over 8}\left(h_{ij}^{\prime 2}-(\partial
h_{ij})^2\right),
\end{eqnarray}
\begin{eqnarray}\label{eq:int-action-massless}
{S}_{\psi{\bar \psi}h}^{(3)} & = & \int d^4x {a^2\over 2}T^{ij}
h_{ij}\nonumber\\
& = & i\int d^4x~\frac{
a^2}{8}\left(\bar{\psi}\gamma^i(\partial^j\psi)+\bar{\psi}\gamma^j(\partial^i\psi)
 -(\partial^i\bar{\psi})\gamma^j\psi-(\partial^j\bar{\psi})\gamma^i\psi\right)h_{ij},
\end{eqnarray}
respectively. Then the equation of motion for $h_{ij}$, sourced by
Dirac field at second-order, can be expressed in conformal time as
\begin{eqnarray}
 h''_{ij}+2\frac{a'}{a}h'_{ij}-\partial^2h_{ij}=\frac{2a}{M_P^2} T_{ij}^{TT},\label{h_ij}
\end{eqnarray}
where $T_{ij}^{TT}=\Pi_{ij}^{~~lm}T_{lm} $, and $T_{ij}$ is
defined in Eq.(\ref{eq:int-action-massless}), and the
TT-projection tensor $\Pi_{ij}^{~~lm}\equiv
P_i^mP_j^l-\frac{1}{2}P_{ij}P^{lm}$, while $P_{ij}$ is the
projection operator defined by
\begin{eqnarray}
 P_{ij}=\delta_{ij}-\frac{k_ik_j}{|\mathbf{k}|^2},
\end{eqnarray}
where $k_i/|\mathbf{k}|$ denotes the direction of the propagation
of a plane wave.
The solution of Eq.(\ref{h_ij}) takes the following form
\begin{eqnarray}
 h_{ij}(\mathbf{k},\tau)=\frac{2}{M_P^2}\int d\tau'G_k(\tau,\tau')a(\tau')T_{ij}^{TT}(\mathbf{k},\tau'),\label{h_ij-1}
\end{eqnarray}
and $G_k(\tau,\tau')$ is the retarded propagator solving the
homogeneous transform of Eq.\eqref{h_ij}. Within a de Sitter
background $a(\tau)=-(H\tau)^{-1}$, $G_k(\tau,\tau')$ reads
\begin{eqnarray}
 G_k(\tau,\tau')=\frac{1}{k^3\tau'^2}[(1+k^2\tau\tau')\sin k(\tau-\tau')+k(\tau'-\tau)\cos k(\tau-\tau')]\Theta(\tau-\tau').\label{greenfunction}
\end{eqnarray}
We will use Eqs. (\ref{h_ij-1}, \ref{greenfunction}) in the
following process.

Using Eq.\eqref{h_ij-1} and plugging in the Fourier transformation
of the Dirac field, we get the tensor spectrum as
\begin{eqnarray}
 &~&\langle h_{ij}(\mathbf{k},\tau)h_{ij}(\mathbf{k}',\tau)\rangle\nonumber\\
 &=&-\frac{1}{4M_P^4}\int d\tau'a(\tau')G_k(\tau,\tau')\int d\tau''a(\tau'')G_k(\tau,\tau'')\int d^3\mathbf{p}d^3\mathbf{p'}
    \Pi_{ijlm}(k)\Pi_{ijno}(k')\nonumber\\
 &~&\langle[\bar{\psi}(\mathbf{p},\tau')\gamma^l\partial^m\psi(\mathbf{k}-\mathbf{p},\tau')
    -\partial^l\bar{\psi}(\mathbf{p},\tau')\gamma^m\psi(\mathbf{k}-\mathbf{p},\tau')+(l\leftrightarrow m)]\nonumber\\
 &~&[\bar{\psi}(\mathbf{p'},\tau'')\gamma^n\partial^o\psi(\mathbf{k'}-\mathbf{p'},\tau'')
    -\partial^n\bar{\psi}(\mathbf{p'},\tau'')\gamma^o\psi(\mathbf{k'}-\mathbf{p'},\tau'')+(n\leftrightarrow o)]\rangle.\label{powerspectrum0}
\end{eqnarray}
To proceed we calculate
\begin{eqnarray}
 &~& a^3(\tau')a^3(\tau'')\sum_{s,s'}\mbox{tr}[\gamma^m X_{\mathbf{p}',s'}(\tau')\bar{X}_{\mathbf{p}',s'}(\tau'')
    \gamma^n W_{\mathbf{p},s}(\tau'')\bar{W}_{\mathbf{p},s}(\tau')]\nonumber\\
 &=&2(\hat{p}'^m\hat{p}^n+\hat{p}'^n\hat{p}^m-\hat{p}'\cdot\hat{p}\delta^{mn})
    (\frac{\Gamma^4(\bar{\mu})}{(2\pi)^8}\left(\frac{-p'\tau'}{2}\right)^{-2ir}\left(\frac{-p\tau''}{2}\right)^{-2ir}
    +\frac{\Gamma^4(\mu)}{(2\pi)^8}\left(\frac{-p'\tau'}{2}\right)^{2ir}\left(\frac{-p\tau''}{2}\right)^{2ir})\nonumber\\
 &~&+2\delta^{mn}\frac{|\Gamma(\bar{\mu})|^4}{(2\pi)^8}(e^{-2\pi r}\tau'^{-2ir}\tau''^{2ir}+e^{2\pi
 r}\tau'^{2ir}\tau''^{-2ir}),
\end{eqnarray}
where $\mu \equiv {1}/{2}-ir$, see Appendix \ref{trace} for
details. Then we have
\begin{eqnarray}
 &~&\langle h_{ij}(\mathbf{k},\tau)h_{ij}(\mathbf{k}',\tau)\rangle\nonumber\\
 &=&\frac{1}{M_P^4}\int d\tau'\frac{G_k(\tau,\tau')}{a^2(\tau')}\int d\tau''\frac{G_k(\tau,\tau'')}{a^2(\tau'')}\int d^3\mathbf{p}
    \delta^3(\mathbf{k}-\mathbf{k'})\nonumber\\
 &~&\{-16A\frac{p^3}{p'}[\frac{1}{2}-(\hat{p}\cdot\hat{k})^2+\frac{1}{2}(\hat{p}\cdot\hat{k})^4]
    -8A\hat{p}\cdot\hat{p}'p^2[1-(\hat{p}\cdot\hat{k})^2]+8Bp^2[1-(\hat{p}\cdot\hat{k})^2]\},\label{powerspectrum1}
\end{eqnarray}
where $A =
\frac{\Gamma^4(\bar{\mu})}{(2\pi)^8}\left(\frac{-p'\tau'}{2}\right)^{-2ir}\left(\frac{-p\tau''}{2}\right)^{-2ir}
    +\frac{\Gamma^4(\mu)}{(2\pi)^8}\left(\frac{-p'\tau'}{2}\right)^{2ir}\left(\frac{-p\tau''}{2}\right)^{2ir}$
and $B = \frac{|\Gamma(\bar{\mu})|^4}{(2\pi)^8}(e^{-2\pi
r}\tau'^{-2ir}\tau''^{2ir}+e^{2\pi r}\tau'^{2ir}\tau''^{-2ir})$.
After marking the angle between $\mathbf{k}$ and $\mathbf{p}$ with
$\theta$ and defining
$x\equiv\frac{|\mathbf{k}-\mathbf{p}|}{k}=\frac{p'}{k},~y\equiv\frac{p}{k}$,
we have
\begin{eqnarray}
 d\cos\theta=-\frac{x}{y}dx,
\end{eqnarray}
since $\cos\theta=\frac{k^2+p^2-p'^2}{2kp}=\frac{1+y^2-x^2}{2y}$,
where $y$ is regarded as constant in the integral before we do the
$y$-integral. When $\cos\theta$ increases monotonically from -1 to
1, or $\theta$ runs backward from $\pi$ to 0,
$x=\sqrt{1+y^2-2y\cos\theta}$ decreases monotonically from $1+y$
to $|1-y|$. Exchanging the $\theta$ and $p$ integrals, we get
\begin{eqnarray}
 &~&\langle h_{ij}(\mathbf{k},\tau)h_{ij}(\mathbf{k}',\tau)\rangle\nonumber\\
 &=&\frac{\delta^3(\mathbf{k}-\mathbf{k'})2\pi k^5}{M_P^4}\int_0^\infty dy\int^{1+y}_{|1-y|}dxxy^3\nonumber\\
 &~&\Big\{[-16\frac{y}{x}(\frac{1}{2}-\cos^2\theta+\frac{1}{2}\cos^4\theta)-8\frac{1-x^2-y^2}{2xy}(1-\cos^2\theta)]\nonumber\\
 &~&\times \Big[\frac{\Gamma^4(\bar{\mu})}{(2\pi)^8}\left(\frac{pp'}{4}\right)^{-2ir}[\int d\tau'\frac{G_k(\tau,\tau')}{a^2(\tau')}\tau'^{-2ir}]^2
    +\frac{\Gamma^4(\mu)}{(2\pi)^8}\left(\frac{pp'}{4}\right)^{2ir}[\int d\tau'\frac{G_k(\tau,\tau')}{a^2(\tau')}\tau'^{2ir}]^2\Big]\nonumber\\
 &~&+8(1-\cos^2\theta)
    \frac{|\Gamma(\bar{\mu})|^4}{(2\pi)^8}\Big[e^{-2\pi r}\int d\tau'\frac{G_k(\tau,\tau')}{a^2(\tau')}\tau'^{-2ir}
    \int d\tau''\frac{G_k(\tau,\tau'')}{a^2(\tau'')}\tau''^{2ir}\nonumber\\
 &~&+e^{2\pi r}\int d\tau'\frac{G_k(\tau,\tau')}{a^2(\tau')}\tau'^{2ir}
    \int d\tau''\frac{G_k(\tau,\tau'')}{a^2(\tau'')}\tau''^{-2ir}\Big]\Big\}.\label{powerspectrum2}
\end{eqnarray}
Note that for $\tau=0$,
\begin{eqnarray}
 G_k(0,\tau')=\frac{1}{k^3\tau'^2}[-\sin k\tau'+k\tau'\cos k\tau'].
\end{eqnarray}
Now, we take both time integrals to cover the range $-\infty\to
0$. To do the time integrals, we combine $k$ and $\tau',~ \tau''$
to form new variables $z'\equiv k \tau',~z''\equiv k \tau''$. Then
we have
\begin{eqnarray}
 &~&\langle h_{ij}(\mathbf{k})h_{ij}(\mathbf{k}')\rangle\nonumber\\
 &=&\frac{\delta^3(\mathbf{k}-\mathbf{k'})2\pi H^4}{M_P^4k^3}\int_0^\infty dy\int^{1+y}_{|1-y|}dxxy^3\nonumber\\
 &~&\Big\{[-16\frac{y}{x}(\frac{1}{2}-\cos^2\theta+\frac{1}{2}\cos^4\theta)-8\frac{1-x^2-y^2}{2xy}(1-\cos^2\theta)]\nonumber\\
 &~&\times \Big[\frac{\Gamma^4(\bar{\mu})}{(2\pi)^8}\left(\frac{xy}{4}\right)^{-2ir}[\int dz' z'^{-2ir}(-\sin z'+z'\cos z')]^2
    +c.c\Big]\nonumber\\
 &~&+8(1-\cos^2\theta)
    \frac{|\Gamma(\bar{\mu})|^4}{(2\pi)^8}\Big[e^{-2\pi r}\int dz' z'^{-2ir}(-\sin z'+z'\cos z'])
    \int dz'' z''^{2ir}(-\sin z''+z''\cos z''])\nonumber\\
 &~&+e^{2\pi r}\int dz' z'^{2ir}(-\sin z'+z'\cos z'])\int dz'' z''^{-2ir}(-\sin z''+z''\cos z''])\Big]\Big\},\label{powerspectrum3}
\end{eqnarray}
where we have used $a(\tau)H=-\frac{1}{\tau}$. Integrating over time and momentum, we get
\begin{eqnarray}
 &~&\langle h_{ij}(\mathbf{k})h_{ij}(\mathbf{k}')\rangle\nonumber\\
 &=&\frac{\delta^3(\mathbf{k}-\mathbf{k'})H^4}{(2\pi)^7M_P^4k^3} Re\Big\{\frac{2^{8ir}\sqrt{\pi}(2i+r)\Gamma(2-2ir)\Gamma^4(\frac{1}{2}+ir)\Gamma^2(-2ir)}{(i+2r)^2\Gamma(\frac{7}{2}-2ir)}
 [4r(i+r)\cosh(\pi r)]^2\mathcal{Z}\Big\},\nonumber\\\label{powerspectrum4}
\end{eqnarray}
where
\begin{eqnarray}
 \mathcal{Z}\equiv\frac{i+r}{-5+2ir}+\frac{{\rm Exp}\{-2\pi r[189+3r(-187i+6r(-41+22ir+4r^2))]\}} {(-9+2ir)(3i+2r)(5i+2r)(7i+2r)}
\end{eqnarray}
During the integration we encountered a lot of incomplete beta
function $B(\Lambda;a,b)$ and the hypergeometric function
$_2F_1(a,b;c;\Lambda)$, where $\Lambda$ is a physical cut-off, and
\begin{eqnarray}
 B(z;a,b)=z^a[\frac{1}{a}+\frac{(1-b)z}{1+a}+O(z^2,z^3...)],\\
 _2F_1(a,b;c;z)=1+\frac{abz}{c}+O(z^2,z^3...).
\end{eqnarray}
We may omit those terms $\sim \Lambda^n, n\geq1$ which should be
renormalized away. Expect these, we do not require any
approximation relevant with $r$. Thus Eq.\eqref{powerspectrum4} is
an analytical result exactly for arbitrary $r$, i.e. $m_{\psi}/H$.

\begin{figure}[htbp]
\centering
\begin{minipage}[t]{0.49\textwidth}
    \centering
    \includegraphics[width=0.85\textwidth]{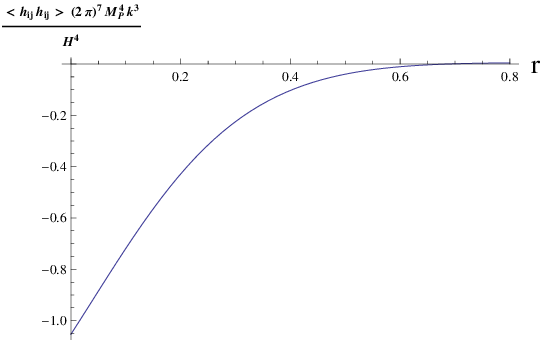}\\
    (a)
\end{minipage}
\begin{minipage}[t]{0.49\textwidth}
    \centering
    \includegraphics[width=0.85\textwidth]{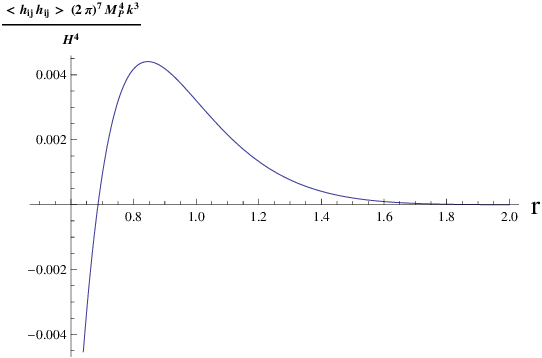}\\
    (b)
\end{minipage}\\
\begin{minipage}[t]{0.49\textwidth}
    \centering
    \includegraphics[width=0.85\textwidth]{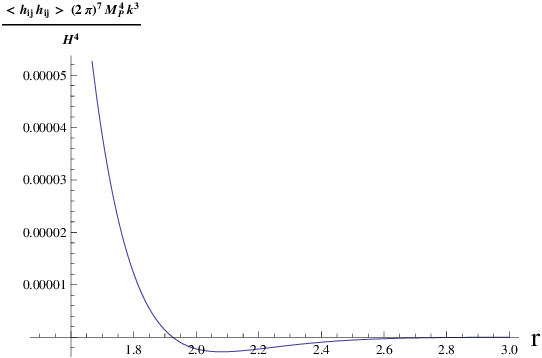}\\
    (c)
\end{minipage}
\begin{minipage}[t]{0.49\textwidth}
    \centering
    \includegraphics[width=0.85\textwidth]{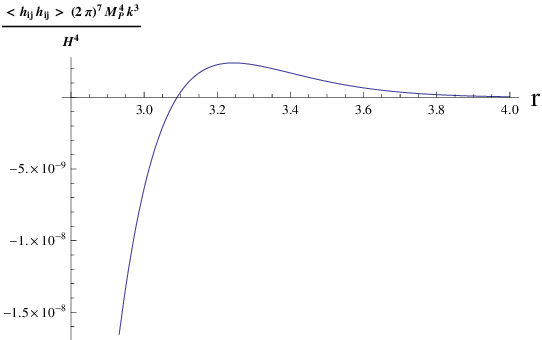}\\
    (d)
\end{minipage}
\caption{The $r$-dependence of Eq.\eqref{powerspectrum4}. We see
that when $r\gtrsim 1$, the Dirac field may contribute negatively
or positively to the power spectrum of gravitational wave,
depending on the value of $r$, i.e. $m_\psi/H$.}
\label{Fig-r-dependence}
\end{figure}

The result involves some Gamma functions written in term of $r$.
However, we can plot its $r$-dependence out easily. In
Fig.\eqref{Fig-r-dependence}, we plot the $Re\{...\}$ part of
Eq.\eqref{powerspectrum4}.
From Fig.\eqref{Fig-r-dependence}, we see that the $r$-dependent
part is $\lesssim 1$, so the contribution of Dirac field to the
gravitational wave power spectrum is $\simeq H^4/M_P^4$, which is
scale-invariant. During inflation, the power spectrum of
gravitational wave is $\simeq H^2/M_P^2$. Thus the contribution of
spinor field is suppressed by the factor $H^2/M_P^2$.
Additionally, from Fig.\eqref{Fig-r-dependence}, we also observe
that when $m_\psi\gtrsim H$, the $r$-dependent part has a
nontrivial behavior, i.e. the Dirac field with different mass may
contribute negatively or positively to the gravitational power
spectrum which manifests itself from the negative or positive
fermion loop correction.

The small mass limit $r\to0$ of Eq.\eqref{powerspectrum4} is
interesting, since this case provides a maximal correction, which
is
\begin{eqnarray}
 \langle h_{ij}(\mathbf{k})h_{ij}(\mathbf{k}')\rangle|_{r\to0}
 =\frac{\delta^3(\mathbf{k}-\mathbf{k'})H^4}{(2\pi)^7M_P^4k^3}\left(-\frac{8\pi^2}{75}\right).\label{powerspectrum5}
\end{eqnarray}

For a check, we follow Ref.\cite{Senatore:2009cf}, regulate the
momentum integral by momentum cut-off, and replace the momentum
integral and time integral as
\begin{eqnarray}
 \int d^3k\to\int^{\Lambda a(t1)}d^3k,  \int_{-\infty}^0\tau''\int_{-\infty}^0\tau'\to\int_{-\infty}^0 d \tau''\int_{-\infty}^{\tau''(1+\frac{H}{\Lambda})}d\tau',
\end{eqnarray}
where $\tau'$ is the earlier one in time and $\Lambda$ is a
physical cut-off. Since the associative cut-off in comoving
momentum is time dependent, therefore we first integrate over
momentum and then integrate over time. After taking the small mass
limit $r\to 0$, we have
\begin{eqnarray}
 \langle h_{ij}(\mathbf{k})h_{ij}(\mathbf{k}')\rangle_{r\to0}
 &=&\frac{\delta^3(\mathbf{k}-\mathbf{k'})\Lambda^5}{480\pi^4k^3HM_P^4(1+\frac{H}{\Lambda})^4}\nonumber\\
 &~&-\frac{\delta^3(\mathbf{k}-\mathbf{k'}) H^4}{(2\pi)^7M_P^4k^3}\Big(
    \frac{64\pi^2(56+44\frac{H}{\Lambda}+18\frac{H^2}{\Lambda^2}+3\frac{H^3}{\Lambda^3})}{75(1+\frac{H}{\Lambda})^4(2+\frac{H}{\Lambda})^3}\nonumber\\
 &~&-\frac{64\pi^2}{75}\frac{\Lambda^3(4+28\frac{H}{\Lambda}+78\frac{H^2}{\Lambda^2}+56\frac{H^3}{\Lambda^3}+44\frac{H^4}{\Lambda^4}
    +18\frac{H^5}{\Lambda^5}+3\frac{H^6}{\Lambda^6})}{H^3(1+\frac{H}{\Lambda})^4(2+\frac{H}{\Lambda})^3}\nonumber\\
 &~&+\frac{64\pi^3\Lambda}{63H}-\frac{64\pi^3\Lambda^3}{105H^3}+\frac{8\pi^3\Lambda^5}{225H^5(1+\frac{H}{\Lambda})^4}\Big),\label{powerspectrum6}
\end{eqnarray}
where the first line on the right-hand side of Eq.
\eqref{powerspectrum6} comes from the $B$-part of Eq.
\eqref{powerspectrum1}, while the remaining parts of Eq.
\eqref{powerspectrum6} come from the $A$-part of Eq.
\eqref{powerspectrum1}. We now leave behind the divergent terms
$\sim \Lambda^n, n\geq1$ assuming that they can be reabsorbed by
counter terms in renormalization process, and we get
\begin{eqnarray}
 \langle h_{ij}(\mathbf{k})h_{ij}(\mathbf{k}')\rangle_{r\to0}
 &\sim&-\frac{\delta^3(\mathbf{k}-\mathbf{k'}) H^4}{(2\pi)^7M_P^4k^3}
    \frac{64\pi^2(56+44\frac{H}{\Lambda}+18\frac{H^2}{\Lambda^2}+3\frac{H^3}{\Lambda^3})}{75(1+\frac{H}{\Lambda})^4(2+\frac{H}{\Lambda})^3}\nonumber\\
 &=&\left.-\frac{\delta^3(\mathbf{k}-\mathbf{k'}) H^4}{(2\pi)^7M_P^4k^3}\frac{64\times56\pi^2}{75}\right|_{\frac{H}{\Lambda}\to0}.\label{powerspectrum7}
\end{eqnarray}
The above result shows the contribution induced by the Dirac
field, in the small mass limit, is of the same order with
Eq. \eqref{powerspectrum5}.

Until now, the impact of quantum
field renormalization on cosmological correlation functions is
controversial. Some authors argued that the influence of
renormalization can significantly change the predictions of
slow-roll inflation for both scalar and tensor power spectra
\cite{Armesto:2007ia, Agullo:2009vq, Agullo:2009zi}.
While some authors debated that the standard expression is valid
and the UV regularisation leads to no difference in the bare power
spectrum \cite{Finelli:2007fr, Durrer:2009ii, Marozzi:2011da, Bastero-Gil:2013nja, Alinea:2015pza}.
Though we focused on the effect of spinor
field on the tensor power spectrum, similar case may also occur.
However, the effect of spinor field is next-to-leading order, so
we expect that available renormalization methods in
literature do not alter our result in order of magnitude.

\section{Conclusion}

Gravity, being the weakest interaction, might provide abundant
information about new physics in the very early universe directly.
Due to the weakness in strength of gravity, it is very hard to
detect the signals of gravitational waves. However, forthcoming
experiments might soon be able to catch some sight of it. This
motivates us to explore possible sources of primordial gravitational
waves.

In this work, we studied the contribution of massive spinor field
during inflation to the gravitational wave power spectrum by using
retarded Green propagator. As an important step, we show the calculation
process of spin sum for massive fermions in APPENDIX A.
We worked out analytically the result
for arbitrary mass $m_{\psi}$, and found that the correction is
scale-invariant and of order $H^4/M_P^4$. Additionally, we also
observe that when the mass of spinor field $m_\psi\gtrsim H$,
depending on the value of $m_\psi$, the spinor field
may contribute negatively or positively to the gravitational power
spectrum.

The result obtained is actually far below the observational level,
which is reasonable since they are of the second order in
perturbation. And of course, we are only limited to the case with
single spinor field, if there are lots of spinor fields, our
result will multiply. In addition, it is also interesting to
consider non-minimal coupling to gravity, e.g. non-minimal
derivative coupling curvaton \cite{Feng:2013pba, Feng:2014tka},
and things might be different. We leave this task to future work.

\textbf{Acknowledgments}

This work is supported by NSFC, No.11222546, and National Basic
Research Program of China, No.2010CB832804.

\appendix

\section{Trace}\label{trace}
Equation Eq. \eqref{powerspectrum0} shows that the calculation of two point correlation function needs to sum over spins of massive fermions at different times. However, massive fermions in curved space are no longer conformally flat, and we can not use the spin sum formula from flat space. Here, we show the calculation process of this spin sum for massive fermions in great details. Despite the differences from the flat space, spin sum always leads to trace of some gamma matrices. the Fourier transformation of the fermion field is
\begin{eqnarray}
 \psi(\mathbf{x},t)&=& \int\mathrm{d}^3p \sum_s \left[ e^{i\mathbf{p}\cdot\mathbf{x}} \alpha_{\mathbf{p},s} X_{\mathbf{p},s}(t) +e^{-i\mathbf{p}\cdot\mathbf{x}} \beta^\dag_{\mathbf{p},s} W_{\mathbf{p},s}(t) \right]\nonumber\\
 &=& a^{-\frac{3}{2}}(t)\int\mathrm{d}^3p \sum_s \left[ e^{i\mathbf{p}\cdot\mathbf{x}} \alpha_{\mathbf{p},s} u_{\mathbf{p},s}(t) +e^{-i\mathbf{p}\cdot\mathbf{x}} \beta^\dag_{\mathbf{p},s} v_{\mathbf{p},s}(t) \right].
\end{eqnarray}
Namely,
\begin{eqnarray}
 X_{\mathbf{p},s}(t)\equiv a^{-\frac{3}{2}}(t)u_{\mathbf{p},s}(t),~W_{\mathbf{p},s}(t)\equiv a^{-\frac{3}{2}}(t)v_{\mathbf{p},s}(t).
\end{eqnarray}
We define
\begin{eqnarray}
 u_{s,\mathbf{p}}(\tau)\equiv\left(\begin{array}{c}
 u_{+,\mathbf{p}}(\tau) \psi_s  \\
 u_{-,\mathbf{p}}(\tau) \psi_s
\end{array}\right);\quad
 v_{s,\mathbf{p}}(\tau)\equiv\left(\begin{array}{c}
 v_{+,\mathbf{p}}(\tau) \psi_{-s}  \\
 v_{-,\mathbf{p}}(\tau) \psi_{-s}
\end{array}\right),
\end{eqnarray}
where $\psi_+=\left(\begin{array}{c}
 1  \\ 0
\end{array}\right)$ and $\psi_-=\left(\begin{array}{c}
 0 \\ 1
\end{array}\right)$ stand for spin up and down, respectively, which make $u_{s,\mathbf{p}}(\tau)$ and $v_{s,\mathbf{p}}(\tau)$ connected by charge conjugation as
$v_{s,\mathbf{p}}(\tau)=C \bar{u}^T_{s,\mathbf{p}}(\tau)=C(u^\dag\gamma^0)^T=C\gamma^0 u^{\dag T}=C\gamma^0 u^*$. Where
\begin{eqnarray}
C=i\gamma^0\gamma^2
 =\left(\begin{array}{cc}
 0 &   i\sigma_2  \\
 i\sigma_2 &   0
 \end{array}\right)
\end{eqnarray}
is charge conjugation matrix. Then
\begin{eqnarray}
 \left(\begin{array}{c}
  v_{+,\mathbf{p}}(\tau) \psi_{-s}  \\
  v_{-,\mathbf{p}}(\tau) \psi_{-s}
 \end{array}\right)
&=&\left(\begin{array}{cc}
 0 &   i\sigma_2  \\
 i\sigma_2 &   0
 \end{array}\right)
 \left(\begin{array}{cc}
 1 &   0  \\
 0 &   -1
 \end{array}\right)
 \left(\begin{array}{c}
 u^*_{+,\mathbf{p}}(\tau) \psi_s  \\
 u^*_{-,\mathbf{p}}(\tau) \psi_s
\end{array}\right)\nonumber\\
&=&\left(\begin{array}{cc}
 0 &   -i\sigma_2  \\
 i\sigma_2 &   0
 \end{array}\right)
 \left(\begin{array}{c}
 u^*_{+,\mathbf{p}}(\tau) \psi_s  \\
 u^*_{-,\mathbf{p}}(\tau) \psi_s
\end{array}\right)\nonumber\\
&=&\left(\begin{array}{c}
 -i\sigma_2u^*_{-,\mathbf{p}}(\tau) \psi_s\\
 i\sigma_2u^*_{+,\mathbf{p}}(\tau) \psi_s
\end{array}\right).
\end{eqnarray}
Therefore
\begin{eqnarray}
 v_{+,\mathbf{p}}(\tau)&=&v_{+,\mathbf{p}}(\tau)\times\mathbf{1}_{2\times2}=\sum_s v_{+,\mathbf{p}}(\tau)\psi_{-s}\psi^T_{-s}\nonumber\\
 &=&\sum_s(-i\sigma_2)u^*_{-,\mathbf{p}}(\tau) \psi_s\psi^T_{-s}=(-i\sigma_2)u^*_{-,\mathbf{p}}(\tau)
  \left(\begin{array}{cc}
   0 &   1  \\
   1 &   0
  \end{array}\right)
 =-i\sigma_2u^*_{-,\mathbf{p}}(\tau)\sigma_1,\\
 v_{-,\mathbf{p}}(\tau)&=&v_{-,\mathbf{p}}(\tau)\times\mathbf{1}_{2\times2}=\sum_s v_{-,\mathbf{p}}(\tau)\psi_{-s}\psi^T_{-s}\nonumber\\
 &=&\sum_si\sigma_2u^*_{+,\mathbf{p}}(\tau) \psi_s\psi^T_{-s}=i\sigma_2u^*_{+,\mathbf{p}}(\tau)
  \left(\begin{array}{cc}
   0 &   1  \\
   1 &   0
  \end{array}\right)
 =i\sigma_2u^*_{+,\mathbf{p}}(\tau)\sigma_1.
\end{eqnarray}
That is we have $v_{\pm,\mathbf{p}}(\tau)=\mp i\sigma_2u^*_{\mp,\mathbf{p}}(\tau)\sigma_1$.
We can now calculate the spin sum.
\begin{eqnarray}
 &~& \gamma^m_{ab}\gamma^n_{cd}\sum_s[\bar{W}_{\mathbf{p},s}(\tau')]_a[W_{\mathbf{p},s}(\tau'')]_d
     \sum_{s'}[X_{\mathbf{p}',s'}(\tau')]_b[\bar{X}_{\mathbf{p}',s'}(\tau'')]_c\nonumber\\
 &=&\frac{1}{a^3(\tau')a^3(\tau'')}\sum_{s,s'}\gamma^m_{ab}[u_{\mathbf{p}',s'}(\tau')\bar{u}_{\mathbf{p}',s'}(\tau'')]_{bc}
    \gamma^n_{cd}[v_{\mathbf{p},s}(\tau'')\bar{v}_{\mathbf{p},s}(\tau')]_{da}\nonumber\\
 &=&\frac{1}{a^3(\tau')a^3(\tau'')}\sum_{s,s'}\mbox{tr}[\gamma^m u_{\mathbf{p}',s'}(\tau')\bar{u}_{\mathbf{p}',s'}(\tau'')
    \gamma^n
    v_{\mathbf{p},s}(\tau'')\bar{v}_{\mathbf{p},s}(\tau')],
\end{eqnarray}
The trace is
\begin{eqnarray}
 &~& \sum_{s,s'}\mbox{tr}[\gamma^m u_{\mathbf{p}',s'}(\tau')\bar{u}_{\mathbf{p}',s'}(\tau'')
    \gamma^n v_{\mathbf{p},s}(\tau'')\bar{v}_{\mathbf{p},s}(\tau')]\nonumber\\
 &=&\sum_{s,s'}\mbox{tr}[\gamma^m u_{\mathbf{p}',s'}(\tau')u^\dag_{\mathbf{p}',s'}(\tau'')\gamma^0
    \gamma^n v_{\mathbf{p},s}(\tau'')v^\dag_{\mathbf{p},s}(\tau')\gamma^0]\nonumber\\
 &=&\sum_{s,s'}\mbox{tr}[
    \left(\begin{array}{cc}
    0 & \sigma_{m} \\
    \sigma_{m}&   0
    \end{array}\right)
    \left(\begin{array}{c}
    u_{+,\mathbf{p}'}(\tau') \psi_{s'}  \\
    u_{-,\mathbf{p}'}(\tau') \psi_{s'}
    \end{array}\right) \left(\psi_{s'}^T u^\dag_{+,\mathbf{p}'}(\tau''),\psi_{s'}^Tu^\dag_{-,\mathbf{p}'}(\tau'')\right)\nonumber\\
 &~&\left(\begin{array}{cc}
    0 & \sigma_{n} \\
    \sigma_{n}&   0
    \end{array}\right)
    \left(\begin{array}{c}
    v_{+,\mathbf{p}}(\tau'') \psi_{-s}  \\
    v_{-,\mathbf{p}}(\tau'') \psi_{-s}
    \end{array}\right) \left(\psi_{-s}^T v^\dag_{+,\mathbf{p}}(\tau'),\psi_{-s}^Tv^\dag_{-,\mathbf{p}}(\tau')\right)]\nonumber\\
 &=&\sum_{s,s'}\mbox{tr}[
    \left(\begin{array}{cc}
    0 & \sigma_{m} \\
    \sigma_{m}&   0
    \end{array}\right)
    \left(\begin{array}{cc}
     u_{+,\mathbf{p}'}(\tau')\psi_{s'} \psi_{s'}^T u^\dag_{+,\mathbf{p}'}(\tau'')
     & u_{+,\mathbf{p}'}(\tau')\psi_{s'}\psi_{s'}^Tu^\dag_{-,\mathbf{p}'}(\tau'') \\
     u_{-,\mathbf{p}'}(\tau') \psi_{s'} \psi_{s'}^T u^\dag_{+,\mathbf{p}'}(\tau'')
     & u_{-,\mathbf{p}'}(\tau') \psi_{s'} \psi_{s'}^Tu^\dag_{-,\mathbf{p}'}(\tau'')
    \end{array}\right)\nonumber\\
 &~&\left(\begin{array}{cc}
    0 & \sigma_{n} \\
    \sigma_{n}&   0
    \end{array}\right)
    \left(\begin{array}{cc}
     v_{+,\mathbf{p}}(\tau'') \psi_{-s} \psi_{-s}^T v^\dag_{+,\mathbf{p}}(\tau')
     & v_{+,\mathbf{p}}(\tau'') \psi_{-s} \psi_{-s}^Tv^\dag_{-,\mathbf{p}}(\tau')\\
     v_{-,\mathbf{p}}(\tau'') \psi_{-s} \psi_{-s}^T v^\dag_{+,\mathbf{p}}(\tau')
     & v_{-,\mathbf{p}}(\tau'') \psi_{-s}\psi_{-s}^Tv^\dag_{-,\mathbf{p}}(\tau')
    \end{array}\right)],
\end{eqnarray}
while
\begin{eqnarray}
 \sum_{s=\pm}\psi_s \psi_s^T=\left(\begin{array}{c}
 0 \\
 1
 \end{array}\right)(0,1)
 +\left(\begin{array}{c}
 1 \\
 0
 \end{array}\right)(1,0)
 =\left(\begin{array}{cc}
 1 &   0  \\
 0 &   1
 \end{array}\right).
\end{eqnarray}
Therefore we have
\begin{eqnarray}
 &~& \sum_{s,s'}\mbox{tr}[\gamma^m u_{\mathbf{p}',s'}(\tau')\bar{u}_{\mathbf{p}',s'}(\tau'')
    \gamma^n v_{\mathbf{p},s}(\tau'')\bar{v}_{\mathbf{p},s}(\tau')]\nonumber\\
 &=&\mbox{tr}[
    \left(\begin{array}{cc}
    0 & \sigma_{m} \\
    \sigma_{m}&   0
    \end{array}\right)
    \left(\begin{array}{cc}
     u_{+,\mathbf{p}'}(\tau') u^\dag_{+,\mathbf{p}'}(\tau'')
     & u_{+,\mathbf{p}'}(\tau') u^\dag_{-,\mathbf{p}'}(\tau'') \\
     u_{-,\mathbf{p}'}(\tau') u^\dag_{+,\mathbf{p}'}(\tau'')
     & u_{-,\mathbf{p}'}(\tau')u^\dag_{-,\mathbf{p}'}(\tau'')
    \end{array}\right)\nonumber\\
 &~&\left(\begin{array}{cc}
    0 & \sigma_{n} \\
    \sigma_{n}&   0
    \end{array}\right)
    \left(\begin{array}{cc}
     v_{+,\mathbf{p}}(\tau'') v^\dag_{+,\mathbf{p}}(\tau')
     & v_{+,\mathbf{p}}(\tau'')  v^\dag_{-,\mathbf{p}}(\tau')\\
     v_{-,\mathbf{p}}(\tau'') v^\dag_{+,\mathbf{p}}(\tau')
     & v_{-,\mathbf{p}}(\tau'') v^\dag_{-,\mathbf{p}}(\tau')
    \end{array}\right)]\nonumber\\
 &=&\mbox{tr}[
    \left(\begin{array}{cc}
     \sigma_m u_{-,\mathbf{p}'}(\tau')  u^\dag_{+,\mathbf{p}'}(\tau'')
     & \sigma_mu_{-,\mathbf{p}'}(\tau')u^\dag_{-,\mathbf{p}'}(\tau'')\\
     \sigma_mu_{+,\mathbf{p}'}(\tau') u^\dag_{+,\mathbf{p}'}(\tau'')
     & \sigma_mu_{+,\mathbf{p}'}(\tau')u^\dag_{-,\mathbf{p}'}(\tau'')
    \end{array}\right)\nonumber\\
 &~&\left(\begin{array}{cc}
     \sigma_n v_{-,\mathbf{p}}(\tau'')  v^\dag_{+,\mathbf{p}}(\tau')
     & \sigma_n v_{-,\mathbf{p}}(\tau'') v^\dag_{-,\mathbf{p}}(\tau')\\
     \sigma_n v_{+,\mathbf{p}}(\tau'') v^\dag_{+,\mathbf{p}}(\tau')
     & \sigma_n v_{+,\mathbf{p}}(\tau'') v^\dag_{-,\mathbf{p}}(\tau')
    \end{array}\right)]\nonumber\\
 &=&\mbox{tr}[\sigma_m u_{-,\mathbf{p}'}(\tau')u^\dag_{+,\mathbf{p}'}(\tau'')\sigma_n v_{-,\mathbf{p}}(\tau'') v^\dag_{+,\mathbf{p}}(\tau')
    +\sigma_mu_{-,\mathbf{p}'}(\tau')u^\dag_{-,\mathbf{p}'}(\tau'')\sigma_n v_{+,\mathbf{p}}(\tau'') v^\dag_{+,\mathbf{p}}(\tau')\nonumber\\
 &~&+\sigma_mu_{+,\mathbf{p}'}(\tau') u^\dag_{+,\mathbf{p}'}(\tau'')\sigma_n v_{-,\mathbf{p}}(\tau'') v^\dag_{-,\mathbf{p}}(\tau')
    +\sigma_mu_{+,\mathbf{p}'}(\tau')u^\dag_{-,\mathbf{p}'}(\tau'')\sigma_n v_{+,\mathbf{p}}(\tau'') v^\dag_{-,\mathbf{p}}(\tau')]\nonumber\\
 &=&\mbox{tr}[
    -\sigma_m u_{-,\mathbf{p}'}(\tau')  u^\dag_{+,\mathbf{p}'}(\tau'')\sigma_n\sigma_2 u^*_{+,\mathbf{p}}(\tau'') u^T_{-,\mathbf{p}}(\tau')\sigma_2^\dag
    +\sigma_mu_{-,\mathbf{p}'}(\tau')u^\dag_{-,\mathbf{p}'}(\tau'')\sigma_n \sigma_2u^*_{-,\mathbf{p}}(\tau'') u^T_{-,\mathbf{p}}(\tau')\sigma_2^\dag\nonumber\\
 &~&+\sigma_mu_{+,\mathbf{p}'}(\tau') u^\dag_{+,\mathbf{p}'}(\tau'')\sigma_n \sigma_2u^*_{+,\mathbf{p}}(\tau'') u^T_{+,\mathbf{p}}(\tau')\sigma_2^\dag
    -\sigma_mu_{+,\mathbf{p}'}(\tau') u^\dag_{-,\mathbf{p}'}(\tau'')\sigma_n \sigma_2u^*_{-,\mathbf{p}}(\tau'') u^T_{+,\mathbf{p}}(\tau')\sigma_2^\dag],\nonumber\\
\end{eqnarray}
where we have used $v_{\pm,\mathbf{p}}(\tau)=\mp
i\sigma_2u^*_{\mp,\mathbf{p}}(\tau)\sigma_1$. The solutions are
\begin{eqnarray}
 u_{s,\mathbf{p}}(\tau)\equiv\left(\begin{array}{c}
 u_{\mu,\mathbf{p}}(\tau)\times \psi_s  \\
 (\hat{p}\cdot\mathbf{\sigma})u_{\bar{\mu},\mathbf{p}}(\tau)\times \psi_s
 \end{array}\right),
\end{eqnarray}
with
\begin{eqnarray}
 &~&u_{+,\mathbf{p}'}(\tau')=\mathbf{1}_{2\times2}u_{\mu,\mathbf{p}}(\tau),\\
 &~&u_{+,\mathbf{p}'}(\tau')=(\hat{p}\cdot\mathbf{\sigma})u_{\bar{\mu},\mathbf{p}}(\tau),
\end{eqnarray}
where $\mu\equiv\frac{1}{2}-ir,\quad\bar{\mu}\equiv\frac{1}{2}+ir$ and
\begin{eqnarray}
u_\mu(x)=\frac{i\sqrt{\pi x}}{2(2\pi)^{3/2}}e^{\frac{\pi r}{2}}H_\mu^{(1)}(x),\\
u_{\bar{\mu}}(x)=\frac{i\sqrt{\pi x}}{2(2\pi)^{3/2}}e^{-\frac{\pi r}{2}}H_{\bar{\mu}}^{(1)}(x).
\end{eqnarray}
Hankel function goes as
\begin{eqnarray}
 H_\beta^{(1)}\simeq-\frac{i\Gamma(\beta)}{\pi}\left(\frac{x}{2}\right)^{-\beta},\quad x\ll1,\beta>0,
\end{eqnarray}
in small $x$ limit. Therefore in this situation, the mode functions behave like
\begin{eqnarray}
u_\mu(x)\simeq\frac{\Gamma(\mu)e^{\frac{\pi r}{2}}}{(2\pi)^2}\left(\frac{x}{2}\right)^{ir},\\
u_{\bar{\mu}}(x)\simeq\frac{\Gamma(\bar{\mu})e^{-\frac{\pi r}{2}}}{(2\pi)^2}\left(\frac{x}{2}\right)^{-ir}.
\end{eqnarray}
Then we have
\begin{eqnarray}
 &~& \sum_{s,s'}\mbox{tr}[\gamma^m u_{\mathbf{p}',s'}(\tau')\bar{u}_{\mathbf{p}',s'}(\tau'')
    \gamma^n v_{\mathbf{p},s}(\tau'')\bar{v}_{\mathbf{p},s}(\tau')]\nonumber\\
 &=&\mbox{tr}[
    -\sigma_m u_{-,\mathbf{p}'}(\tau')  u^\dag_{+,\mathbf{p}'}(\tau'')\sigma_n\sigma_2 u^*_{+,\mathbf{p}}(\tau'') u^T_{-,\mathbf{p}}(\tau')\sigma_2^\dag
    +\sigma_mu_{-,\mathbf{p}'}(\tau')u^\dag_{-,\mathbf{p}'}(\tau'')\sigma_n \sigma_2u^*_{-,\mathbf{p}}(\tau'') u^T_{-,\mathbf{p}}(\tau')\sigma_2^\dag\nonumber\\
 &~&+\sigma_mu_{+,\mathbf{p}'}(\tau') u^\dag_{+,\mathbf{p}'}(\tau'')\sigma_n \sigma_2u^*_{+,\mathbf{p}}(\tau'') u^T_{+,\mathbf{p}}(\tau')\sigma_2^\dag
    -\sigma_mu_{+,\mathbf{p}'}(\tau') u^\dag_{-,\mathbf{p}'}(\tau'')\sigma_n \sigma_2u^*_{-,\mathbf{p}}(\tau'') u^T_{+,\mathbf{p}}(\tau')\sigma_2^\dag]\nonumber\\
 &=&\mbox{tr}[
    -\sigma_m (\hat{p}'\cdot\mathbf{\sigma})u_{\bar{\mu},\mathbf{p}'}(\tau') u^*_{\mu,\mathbf{p}'}(\tau'')\sigma_n\sigma_2 u^*_{\mu,\mathbf{p}}(\tau'') (\hat{p}\cdot\mathbf{\sigma})^Tu_{\bar{\mu},\mathbf{p}}(\tau')\sigma_2^\dag\nonumber\\
 &~&+\sigma_m (\hat{p}'\cdot\mathbf{\sigma})u_{\bar{\mu},\mathbf{p}'}(\tau')(\hat{p}'\cdot\mathbf{\sigma})^\dag u^*_{\bar{\mu},\mathbf{p}'}(\tau'')\sigma_n
    \sigma_2(\hat{p}\cdot\mathbf{\sigma})^*u^*_{\bar{\mu},\mathbf{p}}(\tau'') (\hat{p}\cdot\mathbf{\sigma})^Tu_{\bar{\mu},\mathbf{p}}(\tau')\sigma_2^\dag \nonumber\\
 &~&+\sigma_mu_{\mu,\mathbf{p}'}(\tau') u^*_{\mu,\mathbf{p}'}(\tau'')\sigma_n\sigma_2 u^*_{\mu,\mathbf{p}}(\tau'')
    u_{\mu,\mathbf{p}}(\tau')\sigma_2^\dag\nonumber\\
 &~&-\sigma_mu_{\mu,\mathbf{p}'}(\tau') (\hat{p}'\cdot\mathbf{\sigma})^\dag u^*_{\bar{\mu},\mathbf{p}'}(\tau'')\sigma_n\sigma_2
    (\hat{p}\cdot\mathbf{\sigma})^*u^*_{\bar{\mu},\mathbf{p}}(\tau'')  u_{\mu,\mathbf{p}}(\tau')\sigma_2^\dag]\nonumber\\
 &=&\mbox{tr}[
    -\sigma_m (\hat{p}'\cdot\mathbf{\sigma})\sigma_n \sigma_2(\hat{p}\cdot\mathbf{\sigma})^T\sigma_2^\dag
    u_{\bar{\mu},\mathbf{p}'}(\tau')u^*_{\mu,\mathbf{p}'}(\tau'')u^*_{\mu,\mathbf{p}}(\tau'')u_{\bar{\mu},\mathbf{p}}(\tau')\nonumber\\
 &~&+\sigma_m (\hat{p}'\cdot\mathbf{\sigma})(\hat{p}'\cdot\mathbf{\sigma})^\dag\sigma_n\sigma_2(\hat{p}\cdot\mathbf{\sigma})^*
    (\hat{p}\cdot\mathbf{\sigma})^T\sigma_2^\dag
    u_{\bar{\mu},\mathbf{p}'}(\tau') u^*_{\bar{\mu},\mathbf{p}'}(\tau'')u^*_{\bar{\mu},\mathbf{p}}(\tau'')u_{\bar{\mu},\mathbf{p}}(\tau')\nonumber\\
 &~&+\sigma_m\sigma_n\sigma_2\sigma_2^\dag
    u_{\mu,\mathbf{p}'}(\tau') u^*_{\mu,\mathbf{p}'}(\tau'')u^*_{\mu,\mathbf{p}}(\tau'') u_{\mu,\mathbf{p}}(\tau')\nonumber\\
 &~&-\sigma_m (\hat{p}'\cdot\mathbf{\sigma})^\dag \sigma_n\sigma_2(\hat{p}\cdot\mathbf{\sigma})^*\sigma_2^\dag
    u_{\mu,\mathbf{p}'}(\tau')u^*_{\bar{\mu},\mathbf{p}'}(\tau'')u^*_{\bar{\mu},\mathbf{p}}(\tau'')u_{\mu,\mathbf{p}}(\tau')]\nonumber\\
 &=&\mbox{tr}[
    \sigma_m (\hat{p}'\cdot\mathbf{\sigma})\sigma_n (\hat{p}\cdot\mathbf{\sigma})
    u_{\bar{\mu},\mathbf{p}'}(\tau')u^*_{\mu,\mathbf{p}'}(\tau'')u^*_{\mu,\mathbf{p}}(\tau'')u_{\bar{\mu},\mathbf{p}}(\tau')\nonumber\\
 &~&+\sigma_m \sigma_n
    u_{\bar{\mu},\mathbf{p}'}(\tau') u^*_{\bar{\mu},\mathbf{p}'}(\tau'')u^*_{\bar{\mu},\mathbf{p}}(\tau'')u_{\bar{\mu},\mathbf{p}}(\tau')\nonumber\\
 &~&+\sigma_m\sigma_n
    u_{\mu,\mathbf{p}'}(\tau') u^*_{\mu,\mathbf{p}'}(\tau'')u^*_{\mu,\mathbf{p}}(\tau'') u_{\mu,\mathbf{p}}(\tau')\nonumber\\
 &~&+\sigma_m (\hat{p}'\cdot\mathbf{\sigma})^\dag \sigma_n(\hat{p}\cdot\mathbf{\sigma})
    u_{\mu,\mathbf{p}'}(\tau')u^*_{\bar{\mu},\mathbf{p}'}(\tau'')u^*_{\bar{\mu},\mathbf{p}}(\tau'')u_{\mu,\mathbf{p}}(\tau')],
\end{eqnarray}
where we have use
\begin{eqnarray}
 (\hat{p}\cdot\mathbf{\sigma})(\hat{p}\cdot\mathbf{\sigma})^\dag
 =  \left(\begin{array}{cc}
 \hat{p}_3 & \hat{p}_1-i\hat{p}_2 \\
 \hat{p}_1+i\hat{p}_2 &   -\hat{p}_3
 \end{array}\right)
 \left(\begin{array}{cc}
 \hat{p}_3 & \hat{p}_1-i\hat{p}_2 \\
 \hat{p}_1+i\hat{p}_2 &   -\hat{p}_3
 \end{array}\right)=(\hat{p}_1^2+\hat{p}_2^2+\hat{p}_3^2)\mathbf{1}_{2\times2},
\end{eqnarray}
\begin{eqnarray}
&~&(\hat{p}\cdot\mathbf{\sigma})^*(\hat{p}\cdot\mathbf{\sigma})^T=[(\hat{p}'\cdot\mathbf{\sigma})(\hat{p}'\cdot\mathbf{\sigma})^\dag]^*
    =\mathbf{1}_{2\times2},\\
 &~&\sigma_2(\hat{p}\cdot\mathbf{\sigma})^T\sigma_2^\dag=\sigma_2(\hat{p}\cdot\mathbf{\sigma})^*\sigma_2^\dag=-\hat{p}\cdot\mathbf{\sigma},
\end{eqnarray}
while
\begin{eqnarray}
&~&u_{\bar{\mu},\mathbf{p}'}(\tau')u^*_{\mu,\mathbf{p}'}(\tau'')u^*_{\mu,\mathbf{p}}(\tau'')u_{\bar{\mu},\mathbf{p}}(\tau')\nonumber\\
&=&\frac{\Gamma(\bar{\mu})\Gamma^*(\mu)\Gamma^*(\mu)\Gamma(\bar{\mu})}{(2\pi)^8}
   \left(\frac{-p'\tau'}{2}\right)^{-ir}\left(\frac{-p'\tau''}{2}\right)^{-ir}\left(\frac{-p\tau''}{2}\right)^{-ir}\left(\frac{-p\tau'}{2}\right)^{-ir}
   \nonumber\\
&=&\frac{\Gamma^4(\bar{\mu})}{(2\pi)^8}\left(\frac{-p'\tau'}{2}\right)^{-2ir}\left(\frac{-p\tau''}{2}\right)^{-2ir},\\
&~&u_{\mu,\mathbf{p}'}(\tau')u^*_{\bar{\mu},\mathbf{p}'}(\tau'')u^*_{\bar{\mu},\mathbf{p}}(\tau'')u_{\mu,\mathbf{p}}(\tau')\nonumber\\
&=&\frac{\Gamma(\mu)\Gamma^*(\bar{\mu})\Gamma^*(\bar{\mu})\Gamma(\mu)}{(2\pi)^8}
   \left(\frac{-p'\tau'}{2}\right)^{ir}\left(\frac{-p'\tau''}{2}\right)^{ir}\left(\frac{-p\tau''}{2}\right)^{ir}\left(\frac{-p\tau'}{2}\right)^{ir}
   \nonumber\\
&=&\frac{\Gamma^4(\mu)}{(2\pi)^8}\left(\frac{-p'\tau'}{2}\right)^{2ir}\left(\frac{-p\tau''}{2}\right)^{2ir},
\end{eqnarray}
and
\begin{eqnarray}
&~&u_{\bar{\mu},\mathbf{p}'}(\tau') u^*_{\bar{\mu},\mathbf{p}'}(\tau'')u^*_{\bar{\mu},\mathbf{p}}(\tau'')u_{\bar{\mu},\mathbf{p}}(\tau')\nonumber\\
&=&\frac{\Gamma(\bar{\mu})\Gamma^*(\bar{\mu})\Gamma^*(\bar{\mu})\Gamma(\bar{\mu})}{(2\pi)^8}e^{-2\pi r}
   \left(\frac{-p'\tau'}{2}\right)^{-ir}\left(\frac{-p'\tau''}{2}\right)^{ir}\left(\frac{-p\tau''}{2}\right)^{ir}\left(\frac{-p\tau'}{2}\right)^{-ir}
   \nonumber\\
&=&\frac{|\Gamma(\bar{\mu})|^4}{(2\pi)^8}e^{-2\pi r}\tau'^{-2ir}\tau''^{2ir},\\
&~&u_{\mu,\mathbf{p}'}(\tau') u^*_{\mu,\mathbf{p}'}(\tau'')u^*_{\mu,\mathbf{p}}(\tau'') u_{\mu,\mathbf{p}}(\tau')\nonumber\\
&=&\frac{\Gamma(\mu)\Gamma^*(\mu)\Gamma^*(\mu)\Gamma(\mu)}{(2\pi)^8}e^{2\pi r}
   \left(\frac{-p'\tau'}{2}\right)^{ir}\left(\frac{-p'\tau''}{2}\right)^{-ir}\left(\frac{-p\tau''}{2}\right)^{-ir}\left(\frac{-p\tau'}{2}\right)^{ir}
   \nonumber\\
&=&\frac{|\Gamma(\mu)|^4}{(2\pi)^8}e^{2\pi r}\tau'^{2ir}\tau''^{-2ir}.
\end{eqnarray}
Finally, we get
\begin{eqnarray}
 &~& \sum_{s,s'}\mbox{tr}[\gamma^m u_{\mathbf{p}',s'}(\tau')\bar{u}_{\mathbf{p}',s'}(\tau'')
    \gamma^n v_{\mathbf{p},s}(\tau'')\bar{v}_{\mathbf{p},s}(\tau')]\nonumber\\
 &=&\mbox{tr}[
    \sigma_m (\hat{p}'\cdot\mathbf{\sigma})\sigma_n (\hat{p}\cdot\mathbf{\sigma})
    u_{\bar{\mu},\mathbf{p}'}(\tau')u^*_{\mu,\mathbf{p}'}(\tau'')u^*_{\mu,\mathbf{p}}(\tau'')u_{\bar{\mu},\mathbf{p}}(\tau')\nonumber\\
 &~&+\sigma_m \sigma_n
    u_{\bar{\mu},\mathbf{p}'}(\tau') u^*_{\bar{\mu},\mathbf{p}'}(\tau'')u^*_{\bar{\mu},\mathbf{p}}(\tau'')u_{\bar{\mu},\mathbf{p}}(\tau')\nonumber\\
 &~&+\sigma_m\sigma_n
    u_{\mu,\mathbf{p}'}(\tau') u^*_{\mu,\mathbf{p}'}(\tau'')u^*_{\mu,\mathbf{p}}(\tau'') u_{\mu,\mathbf{p}}(\tau')\nonumber\\
 &~&+\sigma_m (\hat{p}'\cdot\mathbf{\sigma})^\dag \sigma_n(\hat{p}\cdot\mathbf{\sigma})
    u_{\mu,\mathbf{p}'}(\tau')u^*_{\bar{\mu},\mathbf{p}'}(\tau'')u^*_{\bar{\mu},\mathbf{p}}(\tau'')u_{\mu,\mathbf{p}}(\tau')]\nonumber\\
 &=&\mbox{tr}[\sigma_m (\hat{p}'\cdot\mathbf{\sigma})\sigma_n (\hat{p}\cdot\mathbf{\sigma})]
    (\frac{\Gamma^4(\bar{\mu})}{(2\pi)^8}\left(\frac{-p'\tau'}{2}\right)^{-2ir}\left(\frac{-p\tau''}{2}\right)^{-2ir}
    +\frac{\Gamma^4(\mu)}{(2\pi)^8}\left(\frac{-p'\tau'}{2}\right)^{2ir}\left(\frac{-p\tau''}{2}\right)^{2ir})\nonumber\\
 &~&+\mbox{tr}[\sigma_m\sigma_n](\frac{|\Gamma(\bar{\mu})|^4}{(2\pi)^8}e^{-2\pi r}\tau'^{-2ir}\tau''^{2ir}
    +\frac{|\Gamma(\mu)|^4}{(2\pi)^8}e^{2\pi r}\tau'^{2ir}\tau''^{-2ir}).\label{trace1}
\end{eqnarray}

\end{document}